# ASTROMETRIC PLATE REDUCTION WITH ORTHOGONAL FUNCTIONS AND MILLI-ARCSECONDS ACCURACY IN DEEP PROPER MOTION SURVEYS

Devendra Ojha and Olivier Bienaymé

Observatoire de Besançon, BP 1615, F-25010 Besançon Cedex, France

## ABSTRACT

We have been doing a sample survey in UBV photometry and proper motions as part of an investigation of galactic structure and evolution. The programme is based on Schmidt plates (from OCA, Tautenburg, Palomar and ESO Schmidt telescopes) digitized with the MAMA machine (CAI, Paris). The high astrometric quality of the MAMA gives access to micronic accuracy (leading to < 2 mas per year) on proper motions with a 35 years time base. The high proper motion accuracy for faint star probes in wide-areas give access to the properties of star samples out of the solar neighborhood. We have analyzed components of the UVW galactic space motions resulting from the accurate proper motion surveys. The kinematical distribution of F and G-type stars have been probed to distances up to 2.5 kpc above the galactic plane. We have derived the constrain on the structural parameters of the thin and thick disk components of the Galaxy.

# 1 INTRODUCTION

As part of an investigation into stellar evolution and kinematics in the outer regions of the solar neighborhood, 5 regions (at North Galactic Pole, intermediate latitudes and in the galactic plane) were initially selected for our study. UBV Photometric data accurate to 0.05 mag were collected using the combination of ESO, OCA, Tautenburg and Palomar Schmidt plates. Absolute proper motions for all the stars, accurate to 0."2 cen$^{-1}$, were then derived by using the above mentioned plates combination, giving a time base of $\sim 35$ years.

# 2 PLATE TO PLATE TRANSFORMATION

Orthogonal polynomials were used to model the transformation between two plate coordinates. Transformation of coordinates between two catalogue (reference catalogue and $x,y$ of one plate) at the position of $i^{th}$ star write (Bienaymé 1993) :

$$X_i = f_x(x_i, y_i) = \sum_{j=1}^{N} \sum_{k=1}^{n} \alpha_{k,j-k} P_{k,j-k}(x_i, y_i)$$

$$Y_i = f_y(x_i, y_i) = \sum_{j=1}^{N} \sum_{k=1}^{n} \beta_{k,j-k} P_{k,j-k}(x_i, y_i)$$

Where $(x_i, y_i)$ represent the observed coordinates of the star on one of the compared plate and $(X_i, Y_i)$ are the coordinates of the same star after transformation refered to the system of reference plate. $P_{k,j-k}(x_i, y_i)$ is a polynomial in $x_i$ & $y_i$ of order $j$ and $J$ is the order chosen for the transformation.

The distribution of reference stars is used to define an orthogonal system relative to the inner product :

$$(\phi_{k,l}, \phi_{m,n}) = \sum_{i=1}^{N} \phi_{k,l}(x_i, y_i) \phi_{m,n}(x_i, y_i)$$

where N is the number of reference stars. Orthogonality conditions are given by :

$$(\phi_{k,l}, \phi_{m,n}) = 0 \;\; if \;\; (k,l) \neq (m,n)$$

The estimates of the coefficients $\alpha_{k,l}$ and $\beta_{k,l}$ were obtained by minimizing the sum :

$$\sum_{i=1}^{N}(X_i - f_x(x_i y_i))^2$$

$$\sum_{i=1}^{N}(Y_i - f_y(x_i y_i))^2$$

A polynomial of order 6 for the $P_{n,m}$ (28 $\alpha_{n,m}$ or $\beta_{n,m}$ coefficients) gives the best fit with a transform error of less than 0.2 $\mu m$.

The proper motion measurements were performed using relations between plate coordinates only. So the accuracy of these plate to plate relations was limited by the plate centring accuracy. We use all stars of the field to determine the plate relations, but in this particular case it must be assumed that the mean motion of 'standard' objetcs is constant over the field. Proper motions are given by :

$$\mu_x = x_{epoch2} - X(x_{epoch1}, y_{epoch1})$$

$$\mu_y = y_{epoch2} - Y(x_{epoch1}, y_{epoch1})$$

The mean displacement of galaxies and quasars in the reference frame is used to calculate the zero point of the proper motion.

## 3  TRANSFORM ERROR

The transform error or the standard deviation gives an estimate of errors on the determination of $f(x, y)$. It can be evaluated at $x_i, y_i$ for each reference star. It increases with the number of functions used to model the relation between two coordinate catalogues. Figure 1 shows the isolevels of transform error for the 6th order transform between two ESO plates. Both plates were taken at the same epoch, so in this case the transform error is better than 0.2 $\mu m$ over 95% of the field. Figure 2 shows the isolevels of transform error for the 6th order transform between ESO and Palomar plates. In this case the transform error is 0.4 $\mu m$ over most of the field, which is due to the relative motion of the stars during 35 years.

# 4 PROPER MOTION ERROR

The mean error $<\sigma_\mu> = \sqrt{\sigma_{\mu^x}^2 + \sigma_{\mu^y}^2}$ (arcsec per century) in proper motion as a function of V magnitude is shown in figure 3. The overall proper motion accuracy can be deduced from the various sources of errors. Sources of random errors are the plate noise, the digitizing machine, the centering algorithm, and the plate to plate transform. The overall accuracy turns out to be 0."2 cen$^{-1}$ for V<17 and 0."3 cen$^{-1}$ for V>17.

# 5 PHOTOMETRIC AND ASTROMETRIC SURVEYS

Following three fields at intermediate latitude were used for the present study :

• field in the direction of galactic anticentre ($l = 167°$, $b = 47°$ [Ojha et al. 1994a], which covers 7.13 square degree field and is complete down to V=18.5, B=20 and U=16.5 mag).

• field in the direction of galactic centre ($l = 3°$, $b = 47°$ [Ojha et al. 1994b], which covers 15.5 square degree field and is complete down to V=18, B=19 and U=18 mag).

• field in the direction of galactic antirotation ($l = 278°$, $b = 47°$ [in preparation], which covers 20.84 square degree field and is complete down to V=18.5, B=19.5 and U=18 mag).

# 6   RESULTS

We have selected a sub-sample of stars in 0.3≤B-V≤0.9 color or 3.5≤$M_V$≤5 magnitude interval (F and G–type stars). The photometric distance of each star has been determined using a $M_V$ and B-V relation. The UVW velocities have been calculated directly from the measured proper motions in $\mu_l$ and $\mu_b$. The algorithm SEM (Stochastic-Estimation-Maximization; Celeux & Diebolt 1986) is used for the deconvolution of the Gaussian mixtures corresponding to different populations. Through SEM we obtain the number of components of the Gaussian mixture (without any assumption on this number), its mean values, dispersions and the percentage of each component with respect to whole sample. Following structural and kinematical parameters were obtained for the thin and thick disks of the Galaxy :

## 6.1   *Structural parameters*

• From the number ratio of the thin and thick disk stars in a pair of direction towards galactic centre (GC) and anticentre (GAC), we deduced the scale lengths $h_R$ of the thin disk and thick disk, which are found to be 2.6±0.1 and 3.4±0.5 kpc, respectively (see figure 4).

• The density laws for stars with 3.5≤$M_V$≤5 as a function of distance above the plane, follow a single exponential with scale height of ∼ 260 pc for 150≤z≤1200 pc, and a second exponential with scale height of ∼ 800 pc for z distances from ∼ 1200 pc to at least 3000 pc (see figure 5). We identify the 260 pc scale height component as a thin disk, and the 800 pc scale height component as a thick disk, with a local normalization of 5–6 % of the disk.

## 6.2   *Kinematical parameters*

• The thin disk population was found with (<U+W>, <V>) = (1±4, −14±2) km/s and velocity dispersions ($\sigma_{U+W}$, $\sigma_V$) = (35±2, 30±1) km/s.

• The thick disk population was found to have a rotational velocity of $V_{rot}$ = 177 km/s and velocity dispersions ($\sigma_U$,$\sigma_V$,$\sigma_W$) = (67,51,42) km/s (Ojha et al. 1994abc). No dependence with z distance is found in the asymmetric drift of the thick disk population (up to z ∼ 3.5 kpc) (figure 6).

# CAPTION TO THE FIGURES

**Figure 1.** Isocontours of the transform error for 6th order transform between two ESO Schmidt plates. Distance between isocontours is 0.05 micron. Inner isocontour is at 0 micron level.

**Figure 2.** Isocontours of the transform error for 6th order transform between ESO and Palomar Schmidt plates. Distance between isocontours is 0.05 micron. Inner isocontour is at 0 micron level.

**Figure 3.** The mean error $<\sigma_\mu> = \sqrt{\sigma_{\mu^x}^2 + \sigma_{\mu^y}^2}$ (arcsec per century) in proper motion as a function of V magnitude.

**Figure 4.** The observed number of thin and thick disk stars as a function of $r$–distances obtained from SEM algorithm from the two data sets in 0.3$\leq$B-V$\leq$0.9 color interval.

**Figure 5.** The density distribution for stars with 3.5$\leq$M$_V$$\leq$5 as a function of distance above the galactic plane. On this scale, a continuum line represents the sum of two best fitted exponentials. The fitted lines correspond to exponentials with scale height $\sim$ 260 pc and $\sim$ 800 pc and correspond to the 'thin disk' and 'thick disk', respectively.

**Figure 6.** The measured asymmetric drift of the thick disk population plotted as a function of z -distances from the proper motion selected samples (open circle): a– galactic centre survey (Ojha et al. 1994b); (open square): b–anticentre field (Ojha et al. 1994a); (open diamond): c– North Pole (Soubiran 1993); (open triangle): d–galactic antirotation field. Solid line (e) represents Majewski's (1992) survey for only *intermediate population* stars. The dotted (f), dashed (g), dashed-dotted (h) and dotted-dashed-dashed (i) lines represent galactic centre survey, anticentre field, north pole and galactic antirotation field, respectively with *no separation* between the three populations. Vertical bars indicate the error $\frac{\sigma_V}{\sqrt{N}}$ in V velocity. Where N is the number of stars in each distance bin.